\documentclass[iop]{emulateapj}

\slugcomment{}

\shorttitle{Medium-resolution spectroscopy along the Pal 5 stream}
\shortauthors{Ishigaki et al.}

\usepackage{color}
\usepackage{enumerate}
\usepackage{amsmath}

\begin{document}


\title{Line-of-sight velocity and metallicity measurements of the Palomar 5 tidal stream}


\author{M. N. Ishigaki\altaffilmark{1}}
\affil{Kavli Institute for the Physics and Mathematics of 
the Universe (WPI), The University of Tokyo, Kashiwa, Chiba 277-8583, Japan}

\author{N. Hwang\altaffilmark{2}}
\affil{Korea Astronomy and Space Science Institute, 776 Daedeokdae-Ro 
Yuseong-Gu, Daejeon 34055, Korea}

\author{M. Chiba\altaffilmark{3}}
\affil{Astronomical Institute, Tohoku University, Aoba-ku, 
Sendai 980-8578, Japan}

\and

\author{W. Aoki\altaffilmark{4}}
\affil{National Astronomical Observatory of Japan, 2-21-1 Osawa,
  Mitaka-shi, Tokyo 181-8588, Japan}
  



\begin{abstract}
  We present Subaru/FOCAS and Keck/DEIMOS medium-resolution spectroscopy of a
  tidally disrupting Milky Way (MW) globular cluster Palomar 5 and its
  tidal stream. The observed fields are located to cover
  an angular extent of $\sim 17$\arcdeg along the stream, providing an
  opportunity to investigate a trend in line-of-sight velocities ($V_{\rm los}$)
  along the stream, which is essential to constrain its orbit
  and underlying gravitational potential of the Milky Way's dark matter halo. 
  A spectral fitting technique is applied to the observed spectra
  to obtain stellar parameters
 and metallicities ([Fe/H]) of the target stars. 
 The 19 stars most likely belonging to the central Pal 5 cluster
have a mean $V_{\rm los}$ of $-58.1\pm 0.7$ km s$^{-1}$ 
and metallicity [Fe/H]$=-1.35\pm 0.06$ dex, both of which are in good
agreement with those derived in previous high-resolution spectroscopic
studies. Assuming that the stream stars have the same [Fe/H]
as the progenitor cluster, the derived [Fe/H] and $V_{\rm los}$
values are used to estimate the possible $V_{\rm los}$ range
of the member stars at each location along the stream. Because of the heavy
contamination of the field MW stars, estimated $V_{\rm los}$ range
depends on prior assumptions about the stream's $V_{\rm los}$,
which highlights the importance of
more definitely identifying the member stars using proper motion and
chemical abundances to obtain an unbiased information of $V_{\rm los}$
in the outer part of the Pal 5 stream.
The models for the gravitational potential
of the MW's dark matter halo that are compatible with the estimated $V_{\rm los}$ range are discussed. 

\end{abstract}


\keywords{Galaxy: globular clusters: individual: Palomar 5 -- stars: abundances -- stars: kinematics and dynamics}



\section{Introduction}

A stellar tidal stream associated with a globular cluster Palomar 5 (Pal 5)
is considered as one of the sensitive probes of both the global structure and
the substructures of the dark matter halo of the Milky Way (MW) Galaxy
\citep[][and reference there in]{kupper15}.
While streams are now known to be ubiquitous in the MW,
thanks to the recent wide-field photometric surveys,
\citep[e.g.][]{ibata94,grillmair06a,grillmair06b,belokurov07,bonaca12,bernard14}, Pal 5 stream is considered as one of the ideal objects
as a tracer 
of the MW's gravitational potential in the outer stellar halo in
many respects.
Most importantly, the Pal 5 stream has a long and thin morphology
\citep{odenkirchen03,grillmair06c},
which is useful to reconstruct the progenitor's orbit.
\citep[e.g.][]{eyre09,koposov10}.
 Another important property of the Pal 5 stream is that its progenitor
is clearly identified as a globular cluster in the MW halo
at a distance of $\sim 23$ kpc from the Sun. This 
offers an opportunity to investigate mechanisms of tidal disruption
and resulting formation of tidal streams in great detail
\citep[e.g.][]{dehnen04,kupper12,bovy14}. 
It has also been suggested that stellar density fluctuations
along a thin tidal stream like the Pal 5 stream are
the signature of past interactions with numerous
invisible dark matter subhalos
expected to present according to the cold dark matter model
\citep[e.g.][]{yoon11,carlberg12}. Detailed analyses of possible signatures
of the density fluctuation along the Pal 5 stream have
provided implications on the degree of substructures
in the MW's dark matter halo \citep{carlberg12,ibata15,ngan15}.

In order to make a stringent constraint on the structure/substructures
of the MW's dark matter halo using stellar streams, information on
kinematics (i.e. line-of-sight velocities and
proper motions) is crucial. While the kinematic properties of the
progenitor Pal 5 cluster and its neighboring tidal tails have been reported,  
\citep{odenkirchen02,fritz15, odenkirchen09},
little is known on the 
kinematics of the outer part of the stream, which is crucial for
making better constraints on the dark matter halo structure.

One important difficulty in obtaining kinematics
of the stream comes from the fact that it is difficult to unambiguously
identify individual stars belonging
to the stream \citep{odenkirchen09,kuzma15}. Since the progenitor cluster
has relatively low mass \citep[$\sim 6\times 10^3 M_{\odot}$,][]{odenkirchen02}
even at its earlier times \citep[at most $\sim 10$ times more massive than
  present days,][]{odenkirchen03,dehnen04}, stars are sparsely populated on the tidal stream,
especially for the red giant branch
(RGB) stars that are bright enough for spectroscopic
observations \citep{koch04}.

\citet{odenkirchen09} used the line profile of the Mg b triplet
feature around 5180 {\AA} based on a high-resolution
spectroscopy of photometrically pre-selected
stars to separate RGB stars from
contaminating nearby dwarf stars identifying 17 stars
likely belonging to the stream
over an angular extent of $\sim 8.5$\arcdeg. \citet{kuzma15} 
identified the member stars over a larger angular extent
of $\sim 20$\arcdeg, using 
the equivalent widths of \ion{Ca}{2} triplet absorption line
from modest resolution spectra and assuming that the line-of-sight
velocity of the stream stars to be in a range $-70$ to $-35$ km s$^{-1}$. 

To improve the accuracy of the 
membership identification, it is required to first make use of RGB
stars much fainter than the horizontal branch ($r<18$), for which
neither the Mg b method nor the Ca T EWs method are generally
applicable. It is also important to select the
candidate member stars not heavily relying on the line-of-sight
velocities since those of the stream stars and the progenitor
cluster are not necessarily similar to each other, especially
at the stream's outskirts. It has been demonstrated that when
the underlying gravitational potential is not spherical or
not smooth, the outskirts of the stream
would have larger velocity dispersion than expected from a kinematically
cold stream \citep[e.g.][]{yoon11,bonaca14,pearson15,ngan16}.
Therefore, in order to obtain unbiased estimate of kinematics
of the whole stream, it is desirable to assign membership
using information independent of the line-of-sight velocity,
making use of e.g. stellar surface
chemical composition.

In this paper, we present medium-resolution multi-object spectroscopy 
of the Pal 5 cluster and its tidal stream. Our data covers
an angular extent of $\sim 17$\arcdeg along the stream, which is
suitable to evaluate the line-of-sight velocity gradient along the stream.
A spectral fitting method is applied to estimate metallicity ([Fe/H])
of stars.
The measured line-of-sight velocity and [Fe/H] values are then
used to make inference of the line-of-sight velocity at each
location of the stream, taking into account contamination from field
MW stars. While the present data is not large enough
to make strong constraint on the line-of-sight velocity
on the outskirts of the Pal 5 stream given a heavy contamination
of field stars, the method can be
applied to future spectroscopic surveys with next-generation
multi-object spectrographs mounted on large (8-10m-class)
telescopes.

Section \ref{sec:Obs} describes the target selection and 
our Subaru/FOCAS and Keck/DEIMOS observations for the Pal 5 system. 
Section \ref{sec:analysis} describes the method we used to 
measure $V_{\rm los}$ and stellar atmospheric parameters. 
Section \ref{sec:result} presents the results of the  $V_{\rm los}$ 
and [Fe/H] measurement and how these values are used to
estimate probable ranges of $V_{\rm los}$ along the longitudinal
position in the stream. Section \ref{sec:discussion} 
discusses implications on the Galactic 
potential. 

\section[]{Observation and Data Reduction}
\label{sec:Obs}

\subsection{Target fields}

\begin{figure}
\begin{center}
\includegraphics[width=6cm]{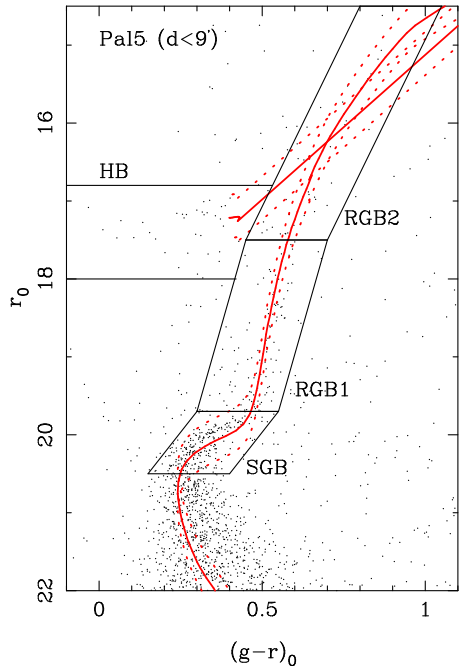}
\caption{Color-magnitude diagram of the Pal 5 central cluster. The four selection boxes used to select the target fields (RGB1, RGB2, SGB, and HB) are
  indicated. Isochrones with an age 11 Gyr and [Fe/H]$=-1.3$ \citep{chen15} are overlaid with varying distances to the Pal 5 cluster, 23.5$\pm 3$ kpc.}
\label{fig:gr_r}
\end{center}
\end{figure}

Target fields along the Pal 5 stream are selected based on the 
stellar number density of the color-magnitude selected sample 
using the SDSS $g$ and $r$ photometry \citep{abazajian09}.
We first defined regions in a
color-magnitude diagram (CMD)
for stars within 9$\arcmin$ from the center of Pal 5. The
four selection boxes, RGB1, RGB2, HB and SGB, together with
theoretical isochrones for an age 11 Gyr, [Fe/H]$=-1.3$ and
distances $23.5\pm 3$ kpc are indicated 
in Figure \ref{fig:gr_r}. 
Then, locations along the Pal 5 stream at which the number density of the CMD- 
selected stars is maximized are identified.   
The resulting positions of the target fields are 
indicated in Figure \ref{fig:fields}.
\citet{grillmair06c} suggests that the distance to  
stream positions vary from 23.2 kpc at the cluster to 23.9 kpc 
at apogalacticon, which corresponds to magnitude difference of 0.06 mag.
This variation does not significantly affect our CMD  
selection methods defined based on stars in the Pal 5 central cluster.

The observations were carried out using 
 the Faint Object Camera and Spectrograph (FOCAS) on the Subaru 
telescope \citep{kashikawa02} and the Deep Imaging Multi-Object Spectrometer 
(DEIMOS; \citet{faber03}) on the Keck II telescope. 
In both spectroscopic observations, the higher priority was given to
stars in the four CMD selection boxes. Most of the other stars
with $r<20.5$ were
also observed in the case of the Keck/DEIMOS observation. 
In the following subsections, we describe details of the 
observations and the line-of-sight velocity measurements.

\subsection{Subaru/FOCAS observation}
\label{sec:focas}

\begin{figure*}
\begin{center}
   \includegraphics[width=15cm]{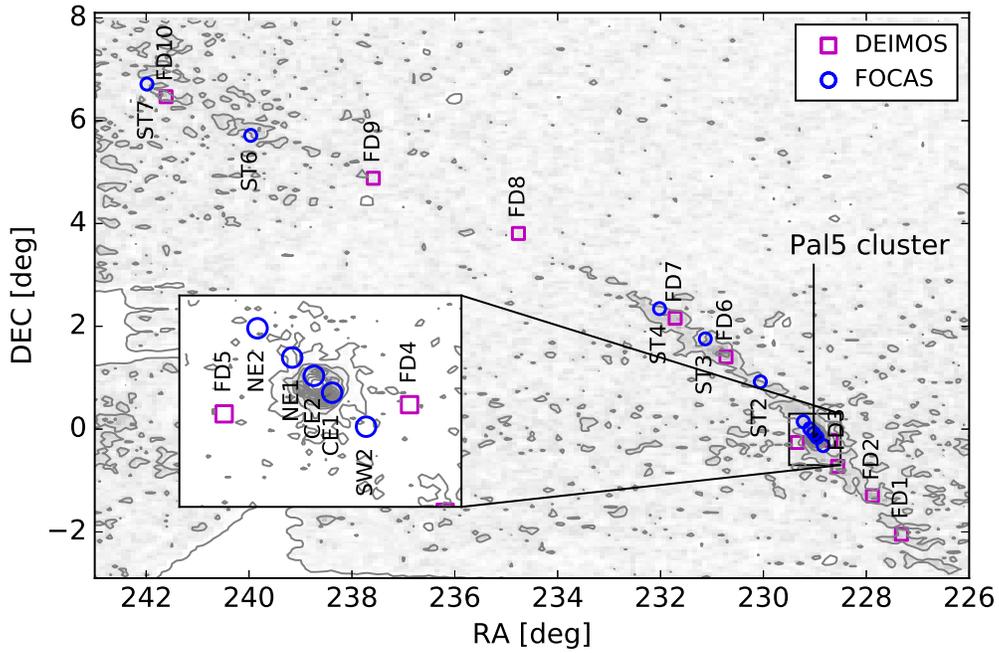}
\caption{Positions of the target fields marked on the stellar number density diagrams constructed with stars selected in Figure \ref{fig:gr_r}. Squares and circles indicate positions of the DEIMOS and FOCAS fields, respectively.}
\label{fig:fields}
\end{center}
\end{figure*}

The Subaru/FOCAS observations were carried out on 
29th and 30th of June 2011. 
The high dispersion VPH grism VPH800, which
covers a wavelength range $\sim 7600-8600$ {\AA}, was used 
with a spectral resolution of $R\sim 7000$.
The central cluster and its outskirts were 
covered with 5 pointings (CE1, CE2, NE1, NE2, and SW2). 
Other pointings were 
targeted at the stream fields (ST2, ST3, ST4, ST6, and ST7).
Locations of the fields are shown in Figure \ref{fig:fields}.
Spectra of metal-poor standard stars with known line-of-sight velocities
(HD 186478, BD--18 5550, BD--17\arcdeg6036, BD+41\arcdeg3931, HD 111721, HD 117936,
and HD 154635) were 
also obtained for the velocity calibration 
with the same grating as the target objects.
For the raw images, distortion calibration, bias subtraction and flat fielding
are performed  with the FOCASRED
package. Then standard IRAF routines are used for the wavelength
calibration using OH skylines and the background sky subtraction.
The derived one-dimensional spectra are then normalized by fitting
a polynomial to the continuum. 

The helio-centric line-of-sight velocities and their associated errors
of the target stars are computed by
cross-correlating spectra of the target stars with those of the
metal-poor standard stars using the IRAF {\it fxcor} task.
Since the field of view (6' diameter) and the number of observed stars
are small for the Subaru/FOCAS observations,
only the data for the central part of Pal 5 cluster 
(CE1 and CE2) are mainly used in the following analyses.
The details of the observations and the number of stars
in each field are summarized in 
Table \ref{tab:focas}.

\begin{deluxetable*}{lccccc}
  \tablecaption{Summary of Subaru/FOCAS observation. \label{tab:focas}}
  \tablewidth{0pt}
\tablehead{
  \colhead{Field name}  & \colhead{RA} & \colhead{DEC} &  \colhead{Date}   & \colhead{Exp. time} & \colhead{$N_{\rm slits}$}
  }
 \startdata
  CE1 &15:15:49.99 & $-$00:09:36.00  &29-6-2011 & 900$\times$4& 21\\
  CE2 & 15:16:05.49 & $-$00:04:48.20&29-6-2011  & 900$\times$4& 26\\
  NE1  &15:16:21.51 & 00:00:07.00 &29-6-2011  & 900$\times$4& 10\\
  NE2 & 15:16:50.48 & 00:08:51.20 & 30-6-2011 & 900$\times$4& 8\\
  SW2 &15:15:20.53 & $-$00:18:21.40 & 29-6-2011 & 900$\times$4& 10\\
  ST2 & 15:20:13.14 &  00:54:54.30 & 30-6-2011 & 900$\times$4& 9\\
  ST3  & 15:24:31.45 &  01:45:57.00 & 30-6-2011& 900$\times$4& 9\\
  ST4 &15:28:05.00 & 02:20:12.50  & 30-6-2011& 900$\times$4& 8\\
  ST6 &15:59:49.21 & 05:42:15.60  & 29-6-2011 & 900$\times$4& 9\\ 
  ST7  &16:07:55.94 & 06:43:25.70  & 30-6-2011& 900$\times$4& 9
  \enddata
\end{deluxetable*}

\subsection{Keck/DEIMOS observation}
\label{sec:deimos}

The Keck/DEIMOS observation was carried out on 12th and 13th of May 2013. 
For the spectroscopic observation, the 1200 lines mm$^{-1}$ grating was
used together with the OG550 filter,
which approximately covers the wavelength range $\sim 6500-9000$ {\AA}.
The slit width of 0''.75 was adopted which 
yields a spectral resolution of $R\sim 6000$. 
Data reduction was performed by the {\it spec2d} DEIMOS DEEP2 reduction 
pipeline \citep{newman13,cooper12}. The reduced one-dimensional
spectra are finally normalized by fitting a polynomial to the continuum.

The helio-centric line-of-sight velocities and their errors are measured
by cross correlating the observed
spectra with a synthetic spectrum for a metal-poor giant star with
$T_{\rm eff}=4500$ K, $\log g=2.5$, [Fe/H]$=-1.5$, and [$\alpha$/Fe]$=0.4$
using the IRAF {\it xcsao} task. 
In the following analysis, objects with spectra with S/N$>10$ and
$V_{\rm los}$ error $< 10$ km s$^{-1}$ are used, which 
amounts to 551 stars in the 10 DEIMOS fields.  
The observing details and the number of stars in each field
are summarized in Table \ref{tab:keck}.

In the DEIMOS fields one star (FD3-008) is
observed in common with \citet{kuzma15} (P1238216). The line-of-sight
velocity for this object obtained in this work is $-60.6\pm 8.7$
km s$^{-1}$, which is in good
agreement with that obtained by \citet{kuzma15} ($-59.8$ km s$^{-1}$).

\begin{deluxetable*}{lccccc}
  \tablecaption{Summary of Keck/DEIMOS observation. \label{tab:keck}}
  \tablewidth{0pt}
  \tablehead{
    \colhead{Field name}  & \colhead{RA} & \colhead{DEC}  &  \colhead{Date (UT)}  & \colhead{Exp. time} & \colhead{$N_{\rm slits}$}
    }
 \startdata
FD1 &             15:09:16.01 & $-$2:02:46.8  & 13-05-2013 &1800$\times$3 & 103 \\
FD2 &              15:11:31.46 & $-$1:17:50.8 & 12-05-2013 &1800$\times$3 & 81 \\
FD3 &              15:14:13.84 & $-$0:43:30.7 & 13-05-2013 &1800$\times$3 & 96 \\ 
FD4 &              15:14:44.11 & $-$0:12:59.7 & 12-05-2013&1800$\times$3 & 78\\
FD5 &              15:17:21.87 &  0:15:36.0 & 13-05-2013 &1800$\times$3 & 71 \\
FD6 &              15:22:54.55 &  1:24:18.9 & 12-05-2013 &1800$\times$3 & 72\\
FD7 &              15:26:51.80 &  2:09:27.1 & 13-05-2013 &1800$\times$3 & 71\\
FD8 &              15:39:03.08 &  3:48:22.2 & 12-05-2013 &1800$\times$3 & 68\\
FD9 &              15:50:20.14 &  4:52:50.8 & 13-05-2013 &1800$\times$3 & 76\\
FD10 &             16:06:27.13 &  6:28:04.2 & 12-05-2013 &1800$\times$2 & 87
\enddata
\end{deluxetable*}

\section{Analysis}
\label{sec:analysis}

After correcting for the redshift adopting the estimated
line-of-sight velocity, stellar atmospheric 
parameters and chemical compositions
($T_{\rm eff}$, $\log g$, [Fe/H], and [$\alpha$/Fe]) are obtained by
the following procedure:

(1) Effective temperatures ($T_{\rm eff}$) are
estimated based on 
the SDSS photometric data by using calibration of \citet{ramirez05}, 
initially assuming [Fe/H]$=-1.5$

(2) Synthetic spectra approximated by the Eq. (\ref{eq:tgmva}) below 
is fitted with a continuum level, $\log g$, [Fe/H] and [$\alpha$/Fe]
as free parameters. The values of $V_{\rm los}$ and $T_{\rm eff}$ are
also allowed to vary within  $\pm V_{\rm los}$ errors and $\pm 50$K,
respectively.

(3) Update the value of [Fe/H] and repeat (1)-(3) until an approximate 
convergence ($\lesssim 0.05$ dex in [Fe/H]) is reached.

Details of the each step are described below.

\subsection{Effective temperature}
The initial values of $T_{\rm eff}$ are obtained based 
on the extinction-corrected SDSS $(g-r)$ color, 
which is transformed to the Johnson $(V-R)$ color 
\citep{jordi06}. Calibrations of \citet{ramirez05}  
for dwarf and giant stars are used to estimate $T_{\rm eff}$ for 
each case initially assuming [Fe/H]$=-1.5$. The $T_{\rm eff}$ values estimated 
from the dwarf and giant calibrations are averaged to obtain 
the $T_{\rm eff}$ used in the following step. 
The [Fe/H] value and dwarf/giant classification
used to estimate the $T_{\rm eff}$ are iteratively 
updated within $\pm 50$ K of the initial value
in subsequent steps to obtain a final value of $T_{\rm eff}$.    

\subsection{$\log g$, [Fe/H], and [$\alpha$/Fe]}
Using the $V_{\rm los}$ and $T_{\rm eff}$ values obtained in the above, 
we fit the observed spectra with synthetic spectra that are expressed 
as functions of line-of-sight velocity, continuum level, stellar atmospheric 
parameters and chemical 
abundances (iron and $\alpha$-elements). In the following subsections, 
we describe the construction of the 
synthetic spectra. 

\subsubsection{Construction of the synthetic spectra}
We first calculate a grid of synthetic spectra 
for $T_{\rm eff}=4000 - 6250$ K, $\log g= 0.2 - 5.0$ dex,
$\xi=0.5-2.5$ km s$^{-1}$,  [Fe/H]$=-2.5 \sim -0.1$, 
[$\alpha$/Fe]$=-0.3 \sim 1.5$, with steps of $250$ K, $0.2$ dex, 
0.5 km s$^{-1}$, 0.2 dex and 0.3 dex, respectively. 
For the spectral synthesis calculation, the LTE abundance analysis
code used in \citet{aoki09} is employed
with the Kurucz model atmosphere \citep{castelli04}
and linelists of \citet{kurucz11}. 
We convolved the theoretical model spectra to make it R~6000 and
  R~7000 to compare with FOCAS and DEIMOS spectra, respectively, when
  the spectral fitting is performed.

\begin{figure}
  \begin{center}
\includegraphics[width=8.5cm]{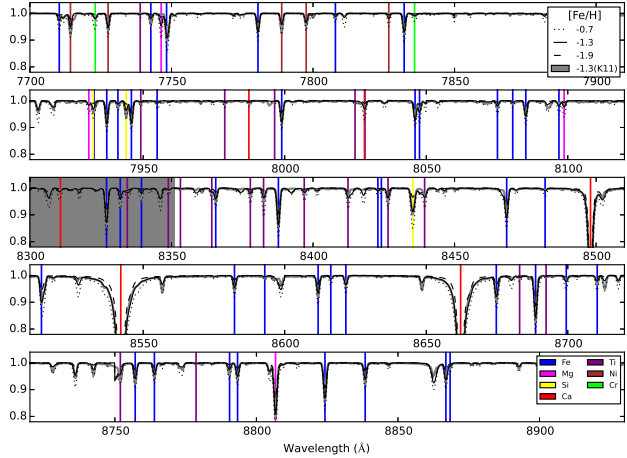}
\caption{Example of the continuum-normalized
  synthetic spectra. The thick gray line shows a spectrum
  from \citet{kirby11}.
  Important spectral features to estimate [Fe/H] are marked
  with vertical lines. The region shaded by gray is not used in the
  spectral fitting because of the heavy contamination with atmospheric
absorption.}
\label{fig:synspecs}
\end{center}
\end{figure}

Example synthetic spectra with three different
metallicities ([Fe/H]$=-0.7, -1.3$ and $-1.9$) and the
other stellar parameters of 
$T_{\rm eff}=5000$K, $\log g=3.0$, and [$\alpha$/Fe]=0.3
are shown in Figure \ref{fig:synspecs}.
The spectral region used in the fitting includes
absorption features of Fe-peak (Cr and Ni) and $\alpha-$ elements
(Mg, Si, Ca, and Ti) that together constrain [Fe/H].
The synthetic spectrum for the case of [Fe/H]$=-1.3$ is
compared with that calculated in \citet{kirby11}
with the same stellar parameters. These are in
good agreement except for molecular features (e.g. CN, C$_{2}$) that are
not included in this work. Since
the molecular features are generally weak compared to the typical
noise level in the spectra used in the present work,
this would have negligible effects on the parameter determination.

The grid of synthetic spectra is then, interpolated to obtain 
formulae  that approximate stellar 
spectra as functions of effective temperature ($\log T_{\rm eff}$), 
surface gravity ($\log g$), micro-turbulent velocity ($\xi$), 
metallicity ([Fe/H]), and $\alpha$-element abundance ([$\alpha$/Fe]) 
at each wavelength point as: 

\begin{eqnarray}
 f(\log T_{\rm eff}, \log g, {\rm [Fe/H]}, \xi, {\rm [}\alpha{\rm /Fe]})= \nonumber\\
a_0+a_1\log T_{\rm eff}+a_2\log g+a_3{\rm [Fe/H]}+a_4\xi+\nonumber\\
a_5{\rm [}\alpha{\rm /Fe]}+a_6(\log T_{\rm eff})^2+a_7(\log g)^2+a_8{\rm [Fe/H]}^2+\nonumber\\
a_9\xi^2+a_{10}{\rm [}\alpha{\rm /Fe]}^2+a_{11}\log T_{\rm eff}\log g+\nonumber\\
a_{12}\log T_{\rm eff}{\rm [Fe/H]}+a_{13}\log T_{\rm eff}\xi+\nonumber\\
a_{14}\log T_{\rm eff}{\rm [}\alpha{\rm /Fe]}+a_{15}\log g{\rm [Fe/H]}+\nonumber\\
a_{16}\log g \xi+a_{17}\log g {\rm [}\alpha{\rm /Fe]}+\nonumber\\
a_{18}{\rm [Fe/H]}\xi+a_{19}{\rm [Fe/H]}[\alpha{\rm /Fe}]+a_{20}\xi{\rm [}\alpha{\rm /Fe]}
\label{eq:tgmva}
\end{eqnarray}

where $a_0$-$a_{20}$ are a set of constants for each wavelength point.

At each wavelength point, we fit the normalized flux values 
with Eq. (\ref{eq:tgmva}) by an IDL routine 
{\it curvefit}, which gives the 
best-fit values of $a_0$-$a_{20}$ at each wavelength.

\subsubsection{Spectral fitting}
Once the values $a_0$-$a_{20}$ have been obtained for the wavelengths in the
range $7700-8900$ \AA, we fit the observed spectra with the synthetic
spectra with a continuum level, $\log g$, 
[Fe/H], [$\alpha$/Fe] as free parameters. In this process, $V_{\rm los}$ 
and $T_{\rm eff}$ are fixed to the values independently obtained 
in the process described above within the uncertainties.
The $\xi$ is also fixed to the 
value obtained by an analytic formula which relates $\log g$ and 
$\xi$ used in Eq. (2) of \citet{kirby09}. 
 We consider the wavelength range of
$7700-8900$ \AA, while available wavelengths vary among the target
stars depending on their slit locations in the FOCAS or DEIMOS masks. 
The wavelength range $8128-8351$ {\AA} is excluded from the fitting
because of the heavy contamination of atmospheric absorption.
Cores of the strong \ion{Ca}{2} triplet
lines within $1.0$ {\AA} are also excluded
since departure from the LTE approximation is expected \citep{kordopatis11}. 

The uncertainties in the $\log g$, [Fe/H] and [$\alpha$/Fe] obtained
by the spectral fitting are evaluated by creating mock spectra
for the stellar parameters $T_{\rm eff}=5000$ K, $\log g=3.0$,
[Fe/H]$=-1.5$, and [$\alpha$/Fe]$=0.3$.
For the each case of S/N$=10$, $30$, and $50$, 200 mock spectra
are created by adding noises obeying a Gaussian
distribution with a $\sigma$ corresponding to the S/N.
In this process, the initial value of $T_{\rm eff}$ is
assumed to be known (5000 K) with
random errors of a Gaussian with $\sigma =100$ K.
A resulting distribution of the $\log g$ estimated from the 200
mock spectra has a mean of 2.85 dex
with a standard deviation of 0.50 dex for the case of S/N$=10$.
For this S/N, the obtained values of [Fe/H] and [$\alpha$/Fe] have means of
$-1.42$ and $0.46$ dex, with dispersions
of 0.14 and 0.17 dex, respectively. The dispersions are
smaller for the higher S/N, while the offsets of up to $\Delta$[Fe/H]$=0.08$ 
dex and $\Delta$[$\alpha$/Fe]$=0.18$ dex
remain even for the highest S/N case. From this exercise,
the internal errors in [Fe/H] of at least $0.14$ dex should be
taken into account. This is much smaller than systematic
uncertainties evaluated in the next section.

\subsubsection{Comparison with high-resolution spectroscopy}

We test the method described above by comparing our estimates of
stellar parameters ($T_{\rm eff}$, $\log g$, [Fe/H] 
and [$\alpha$/Fe]) with those estimated based on high-resolution 
spectroscopy ($R\ge 30000$) in literature. 
In the Subaru/FOCAS observations we took long-slit spectra of two
bright metal-poor stars, HD 111721 and HD 186478, for which
high-resolution spectroscopic analyses are available from
\citet{ishigaki12} and \citet{cayrel04}, respectively.
In the Keck/DEIMOS observations, two bright stars 
FD6-042 and FD6-067 in our sample have been analyzed using
high-resolution spectra by \citet{katz11}.
For the comparison, we used four different
color-$T_{\rm eff}$ calibrations to estimate initial $T_{\rm eff}$
values adopted in our analysis
procedures. The results are summarized in 
Table \ref{tab:standard}.

For the four stars used in the comparison,
the different color-$T_{\rm eff}$ relations can result in
up to $\sim 400$ K differences in the $T_{\rm eff}$. These differences
lead to a large variation in $\log g$ up to $\sim 1$ dex,
which indicates a strong correlation between $T_{\rm eff}$
and $\log g$ in the analyses adopted in this work. 
This can be understood as the wavelength range used in the
fitting includes only a few significant
$\log g$-sensitive absorption lines. One of the 
absorption lines within the fitted wavelength range
sensitive to $\log g$ for cool
($T_{\rm eff}\lesssim 5500$ K) stars is the 
strong \ion{Mg}{1} line at 8807 {\AA}, whose
pressure-broadened wings are expected for higher
$\log g$. This feature is, however, relatively weak
for the lower [Fe/H] star HD 186478, which likely
causes the $>1$ dex discrepancy in $\log g$ depending on
the adopted $T_{\rm eff}$ values.  

On the other hand, the [Fe/H] values do not
largely depend on the adopted color-$T_{\rm eff}$ scales. 
For the $(V-R)$ color with the \citet{ramirez05} scale,
which is used for the remaining target stars,
the obtained [Fe/H] values are 
 in agreement with
those derived with the high-resolution spectroscopy
within 0.25 dex. These comparisons suggest that 
systematic uncertainties in the [Fe/H] estimates
of up to $\sim 0.25$ dex in stars in a range $-2.6<$[Fe/H]$<-0.1$. 
Because of the relatively large uncertainty in the $\log g$ and [$\alpha$/Fe] 
estimates, we only use [Fe/H] to identify
candidate stream stars in the following analyses. 
Precision of the [Fe/H] values is further investigated 
in Section \ref{sec:pal5cen}.

\begin{deluxetable*}{lccccccc}
\tablecaption{Comparison with high-resolution spectroscopy\label{tab:standard}}
\tablewidth{0pt}
\tablehead{
  \colhead{Star name} & \colhead{parameters} & \multicolumn{5}{c}{TW\tablenotemark{a}} & \colhead{HR\tablenotemark{b}}\\
  \colhead{} & \colhead{} & \colhead{$(V-R)$/RM05} & \colhead{$(V-I)$/RM05} & \colhead{$(V-K)$/RM05} & \colhead{$(V-K)$/C10} & \colhead{$(V-K)$/A99} & \colhead{}
}
\startdata
HD 111721 & $T_{\rm eff}$ (K)& 4969 & 5211  & 4993& \nodata & 4979 & 4947 \\
          & $\log g$ & 2.8&  3.6 & 2.9 & \nodata & 2.8 & 2.63 \\
          & [Fe/H] & $-1.27$ & $-1.18$ & $-1.25$ &\nodata& $-1.2$4& -1.33 \\
          & [$\alpha$/Fe] & 0.54 & 0.46 & 0.52 &\nodata& 0.52 & 0.35\\ 
\tableline
HD 186478 & $T_{\rm eff}$ (K)& 4720 &5033 & 4670& \nodata & 4651 & 4700\\
          & $\log g$ & 2.3 &3.3 & 2.2 & \nodata & 2.1 & 1.3\\
          & [Fe/H] & $-2.37$ & $-2.09$ & $-2.43$& \nodata &$-2.44 $ & $-2.59$ \\
          & [$\alpha$/Fe] & 0.47 & 0.18 & 0.51 & \nodata & 0.53 &  0.36\\
\tableline
FD6-042 &  $T_{\rm eff}$ (K)& 6059 & 5614 & 5909 & 5988 &\nodata & 5861\\
 (FM5-46436)     & $\log g$ & 5.0 & 4.9 & 5.0 & 5.0 & \nodata& 4.23\\
& [Fe/H] & $-0.03$ & $-0.16$ & $-0.08$ &$-0.08$ & \nodata & $0.09$\\
& [$\alpha$/Fe] & $0.06$ & $-0.03$ & $-0.01$ &$0.02$ & \nodata & \nodata \\
\tableline
FD6-067 &$T_{\rm eff}$ (K)& 5752 & 5508 & 5908 & 6028 & \nodata & 5892\\
 (FM5-49663)   & $\log g$ & 5.0 & 4.6 & 5.0 & 5.0 & \nodata & 4.06\\
& [Fe/H] & $-0.91$ & $-0.94$ & $-0.83$ & $-0.90$ & \nodata & $-0.72$ \\
& [$\alpha$/Fe] & $0.04$ & $0.09$ & $0.08$ & $0.20$ & \nodata & \nodata 
\enddata
\tablenotetext{a}{The initial $T_{\rm eff}$ values are obtained using different
  color/color-$T_{\rm eff}$ calibrations (RM05: \citet{ramirez05},
  C10: \citet{casagrande10}, and A99: \citet{alonso99}). The
  $V, R, I$ magnitudes are based on
  \citet{zacharias05}, \citet{monet03}, or the SDSS photometry,
  while the $K$ band magnidues are taken from the 2MASS
  catalog \citet{cutri03}.
 Other parameters are obtained by fitting synthetic spectra to observed ones.}
\tablenotetext{b}{The parameters obtained based on high-resolution spectroscopy from literature. HD 111721: \citet{ishigaki12}, HD 186478: \citet{cayrel04}. FD6-042, FD6-067: \citet{katz11}.The [$\alpha$/Fe] values 
given here are obtained by averaging the [Mg/Fe], [Si/Fe], [Ca/Fe], [\ion{Ti}{1}/Fe] and [\ion{Ti}{2}/Fe] values in these literature. }
\end{deluxetable*}

\subsection{Proper motion}
\label{sec:propermotion}
To exclude objects that unlikely belong to the stream, 
proper motion of the target stars are obtained from 
the Initial Gaia Source List (IGSL) \citep{smart13} catalog, when available. 
We use the proper motion data only if errors in $\mu_{\alpha}$ 
and $\mu_{\delta}$ are both smaller than 50 \%.   

If the stream is kinematically cold, stars belonging to the stream
have a proper motion vector approximately aligned with
the projected location of the 
stream and move toward the direction of motion 
of the Pal 5 cluster. 
At each sky position along the stream, we compare proper motion 
vectors of the target stars (transformed to the Galactic coordinate) 
with a vector tangent (positive toward increasing $l \cos b$ and $b$ 
direction) to the stream. The angle ($\theta$) between 
these vectors should be around $\theta \sim 180$\arcdeg or $\cos \theta\sim -1$ 
if the star belongs to the stream.  Therefore, we exclude 
objects with $\cos \theta> -0.5$ from the candidate stream stars.  
We also exclude objects with either $\mu_{l}\cos b>20$ or
$\mu_{b}>20$ mas/yr since the proper motion of the Pal 5
has been reported to be much smaller \citep{fritz15}. 

Since the proper motions are either unavailable or subject to large
uncertainties 
for the majority of the target stars, the proper motion criterion
has only a minor effect 
in the selection of the candidate stream stars
(see Section \ref{sec:selection}).

\section{Results}
\label{sec:result}

\subsection{Velocity and metallicity of the Pal 5 cluster \label{sec:pal5cen}}
 
Among the stars in the two FOCAS fields covering the 
central region of the Pal 5 cluster, CE1 and CE2,
we select the putative member stars 
belonging to the Pal 5 central cluster.  
The selection is made by iteratively 
eliminating three-sigma outliers of the line-of-sight velocities and the
[Fe/H] values from the target stars. 
This process leaves 19 stars that are most likely belonging to the 
central cluster. 
The estimated properties of these stars  are summarized 
  in Table \ref{tab:Pal5center}.

For these Pal 5 cluster stars, we first estimate the
[Fe/H] and [$\alpha$/Fe] treating $\log g$ as a free parameter ($\log g_{\rm spec}$, columns 7-9)
as has been done for the rest of the target stars.
This results in, however, unrealistic values of $\log g$ for some objects such
that the $\log g$ values are not consistent with the distance (23.5 kpc) to
the Pal 5 cluster. In order to check the robustness of the [Fe/H] estimates
against the change in $\log g$, we next estimate the [Fe/H] and [$\alpha$/Fe]
by fixing $\log g$ to the one derived by the standard relation between
an apparent magnitude, an effective temperature and a distance ($\log g_{\rm photo}$, columns 10-12).
As can be seen from the Table \ref{tab:Pal5center}, the spectral fitting method
tend to overestimate $\log g$ by 0.9 dex on average.
Accordingly, the resulting values of [$\alpha$/Fe] in the two methods 
are different by $>0.5$ dex for some of the target stars.
On the other hand, the [Fe/H] values between the two methods remain
similar with a mean difference of [Fe/H]($\log g_{\rm spec}$)-[Fe/H]($\log g_{\rm photo}$)$=0.01$
dex with the standard deviation of 0.18 dex, which is well within the expected
systematic uncertainties. The large uncertainties in $\log g$, therefore,
would not significantly affect the stream membership analyses below 
based solely on the $V_{\rm los}$ and [Fe/H] values.

\begin{figure}
\includegraphics[width=8.5cm]{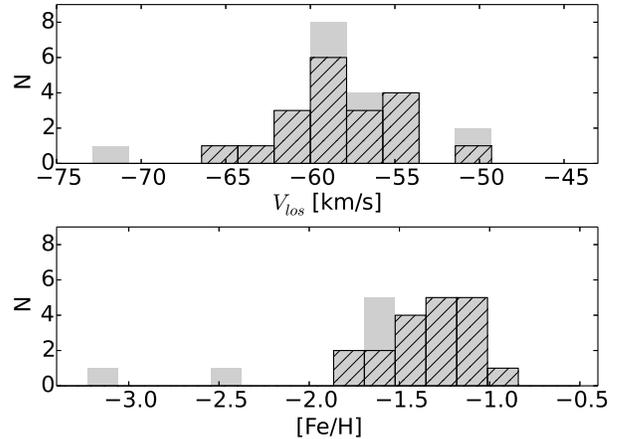}
\caption{Distribution of the line-of-sight velocities (top) and the [Fe/H]
  values (bottom) for
  the stars in the central region of Pal 5. Hatched histograms are for stars
  identified as the Pal 5 cluster members.
  }
\label{fig:vlos_feh}
\end{figure}

Figure \ref{fig:vlos_feh} shows the line-of-sight velocity
and [Fe/H] distributions of stars 
in the CE1 and CE2 fields. The distributions
for the selected member 
stars are shown in the hatched histograms. 
Based on these 19 sample stars, a mean $V_{\rm los}$ of the Pal 5 cluster is 
estimated to be $-58.1\pm 0.7$ km s$^{-1}$, 
which is in good agreement with the value obtained 
by \citet{odenkirchen02} ($-58.7\pm 0.2$ km s$^{-1}$).
The line-of-sight velocity dispersion is estimated to be
$\sigma_{V_{\rm los}}=3.2$ km s$^{-1}$, which is compatible with a typical
uncertainty in the $V_{\rm los}$ measurements in this work. 
This is consistent with a very small intrinsic velocity dispersion of
$<1$ km s$^{-1}$ obtained by \citet{odenkirchen02}.

The mean [Fe/H] value for the 19 stars is $-1.35\pm 0.06$ dex, which is
also in good agreement with 
the mean [Fe/H] value ($\sim -1.3$) obtained from the
high-resolution spectroscopy 
of four red-giant stars in Pal 5 by \citet{smith02}.
A dispersion in the [Fe/H] distribution is $\sigma_{\rm [Fe/H]}=0.25$ dex,
which gives a $1\sigma$ upper limit on any internal metallicity spread. 
Since the high-resolution spectroscopy of the cluster members
reported a very small [Fe/H] spread for Pal 5 \citep{smith02},   
the [Fe/H] dispersion in this work is likely
dominated by precisions in our [Fe/H] measurements.
In the following, we take the value of $\sigma_{\rm [Fe/H]}=0.25$ dex
as a typical value for an uncertainty in our [Fe/H] measurements.

\begin{deluxetable*}{ccccccccccccc}
  \tablecaption{Identified Pal 5 member stars in the FOCAS fields
  \label{tab:Pal5center}}
  \tablewidth{0pt}
  \tablehead{
    \colhead{Field} & \colhead{Star} & \colhead{RA} & \colhead{DEC} & \colhead{$V_{\rm los}$} & \colhead{$T_{\rm eff}$} & \multicolumn{3}{c}{$\log g_{\rm spec}$} & \multicolumn{3}{c}{$\log g_{\rm photo}$}&  \colhead{S/N}\\
    \cline{7-9} \cline{10-12} \\
    \colhead{} & \colhead{} & \colhead{} & \colhead{} & \colhead{} & \colhead{} & \colhead{$\log g$} & \colhead{[Fe/H]} & \colhead{[$\alpha$/Fe]}& \colhead{$\log g$} & \colhead{[Fe/H]} & \colhead{[$\alpha$/Fe]} & \colhead{}\\
    \colhead{} & \colhead{} & \colhead{(deg)} & \colhead{(deg)} & \colhead{(km s$^{-1}$)} & \colhead{(K)} & \colhead{(dex)} & \colhead{(dex)} & \colhead{(dex)} & \colhead{(dex)} & \colhead{(dex)} & \colhead{(dex)} & \colhead{}
  }
  \startdata
  CE1 & s1 & 228.99337 & $-0.14527  $ & $-59.7   $ & 5400 & 4.6 & $-1.36 $ & $0.23  $ & 3.6 & $-1.25$  &  0.50 &  23 \\
  CE1 & s5 & 228.98538 & $-0.14497  $ & $-57.6   $ & 5700 & 5.0 & $-1.00 $ & $-0.29 $ & 3.8 &  $-1.02$ &  0.08  & 12 \\
CE1 & s6 & 229.00683 & $-0.16416  $ & $-60.3   $ & 5270 & 4.6 & $-1.17 $ & $-0.06 $ &3.6 &  $-1.08$  &  0.18 & 21 \\
CE1 & s7 & 229.00375 & $-0.17778  $ & $-56.4   $ & 5218 & 4.9 & $-1.85 $ & $0.25  $ & 3.7 &  $-1.47$ &  0.72 & 11 \\
CE1 & s8 & 228.95590 & $-0.11171  $ & $-55.5   $ & 5464 & 5.0 & $-1.28 $ & $0.12  $ &3.8  & $-1.13$   & 0.50 & 29 \\
CE1 & s11 & 228.97464 & $-0.14002  $ & $-55.0   $ & 5041 & 3.1 & $-1.08 $ & $0.37  $ & 2.8 &  $-1.12$ &  0.40 & 40 \\
CE1 & s13 & 228.96814 & $-0.11985  $ & $-54.6   $ & 4987 & 2.9 & $-1.52 $ & $0.64  $ & 3.1 & $-1.52$ &  0.62 & 24 \\
CE1 & s15 & 228.98008 & $-0.17857  $ & $-51.2   $ & 5076 & 3.7 & $-1.22 $ & $0.50  $ & 3.1 &  $-1.20$ &  0.58 & 20 \\
CE1 & s21 & 229.00542 & $-0.16097  $ & $-62.2   $ & 5502 & 4.3 & $-1.46 $ & $0.26  $ & 2.5 &  $-1.71$ &  0.91 &  14 \\
CE2 & s8 & 229.00142 & $-0.11537  $ & $-59.3   $ & 5927 & 5.0 & $-1.33 $ & $0.20  $ & 3.9 &  $-1.61$  & 1.19 & 13 \\
CE2 & s10 & 229.04567 & $-0.07553  $ & $-59.6   $ & 5910 & 5.0 & $-1.25 $ & $0.13  $ & 3.9 &  $-1.60$ &  1.49 & 22 \\
CE2 & s11 & 229.01102 & $-0.11853  $ & $-58.9   $ & 5389 & 4.5 & $-1.55 $ & $0.03  $ & 3.7&  $-1.51$ &  0.25 & 19 \\
CE2 & s14 & 229.03761 & $-0.10854  $ & $-60.6   $ & 5159 & 1.8 & $-1.67 $ & $0.34  $ & 3.0 &   $-1.46$ &  0.17 & 41 \\
CE2 & s16 & 228.99309 & $-0.10317  $ & $-56.5   $ & 4985 & 3.3 & $-1.07 $ & $0.30  $ & 3.0 & $-1.10$  &  0.34 & 30 \\
CE2 & s17 & 229.06385 & $-0.10081  $ & $-60.8   $ & 5097 & 4.0 & $-1.22 $ & $0.36  $ &2.5 & $-1.29$  &  0.65 &  10 \\
CE2 & s18 & 229.01076 & $-0.09384  $ & $-58.2   $ & 4902 & 2.4 & $-1.86 $ & $0.32  $ & 2.9&  $-1.79$  &  0.27 & 13 \\
CE2 & s22 & 229.02013 & $-0.12284  $ & $-65.3   $ & 5098 & 5.0 & $-1.18 $ & $0.01  $ & 3.3&  $-1.09$  & 0.47& 12 \\
CE2 & s23 & 228.97912 & $-0.08659  $ & $-58.0   $ & 5280 & 5.0 & $-1.14 $ & $0.18  $ & 2.6&  $-1.12$  &  0.68 & 22 \\
CE2 & s24 & 229.03326 & $-0.12798  $ & $-54.6   $ & 5544 & 3.9 & $-1.39 $ & $0.52  $ & 2.5 &  $-1.75$  &  1.47 & 11 
\enddata
\end{deluxetable*}

\subsection{Selection of the candidate stream stars} 
\label{sec:selection}

\begin{figure}
\includegraphics[width=8.5cm]{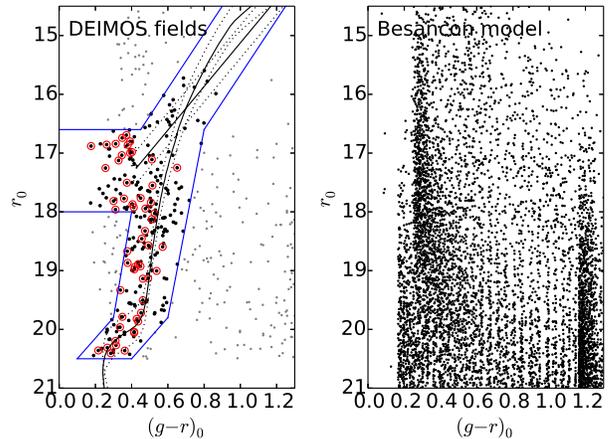}
\caption{{\it Left}: The color-magnitude selection box (solid blue) to select candidate member stars. Gray dots indicate all sample stars observed with DEIMOS. Black dots indicate the sample stars satisfying S/N$>10$, $V_{\rm los}$ error$<10$ km s$^{-1}$ and the color-magnitude selection criterion. Red open circles mark the candidate stream stars (see text in Section \ref{sec:selection}). Isochrones with an age 11 Gyr and [Fe/H]$=-1.3$ \citep{chen15} with varying distances 23.5$\pm 3.0$ kpc are overlaid with black solid and dotted lines. {\it Right}: Simulated Milky Way foreground stars by Besan\c{c}on model.\label{fig:cmd_candidate}}
\end{figure}

Based on the estimated [Fe/H] values, 
together with the available photometry and proper motion data, we select
candidate members of the Pal 5 stream by following criteria: 

\begin{enumerate}
\item[1.] Location in the CMD is close to the locus of the Pal 5's
  giant/subgiant/horizontal branch (an area surrounded by the solid lines
  in  Fig. \ref{fig:cmd_candidate}). The wider color and magnitude
  ranges than those of the cluster stars are adopted to take into account
  a possible distance variation along the stream.
   Adopting this criterion
   leave $248$ candidate stars in the 10 DEIMOS fields.
   
\item[2.] The estimated [Fe/H] value is consistent with
  that of the central cluster ($-1.35$ dex) within $3\sigma$ ($\sim0.75$ dex).
  This reduces the number of candidate stars to 141. 
  
\item[3.]  The proper motions satisfy the criteria described in 
Sec. \ref{sec:propermotion}, if available.
Note that stars with proper motion errors $>50$ \%
are not removed by this criterion. 
As mentioned before, the proper motion is either unavailable or associated
with large uncertainties for the majority of the sample stars. 
As a result, this criterion reduces only
$\sim 8\%$ of the stars selected in the previous steps.
\end{enumerate}

In order to evaluate possible contamination of field MW stars,
we use the Besan\c{c}om model \citep{robin03} to create simulated
catalogs of photometry and kinematics for the field stars within
1 deg$^{2}$ fields centered at the locations of the each DEIMOS
field.

\begin{figure}
\includegraphics[width=8.5cm]{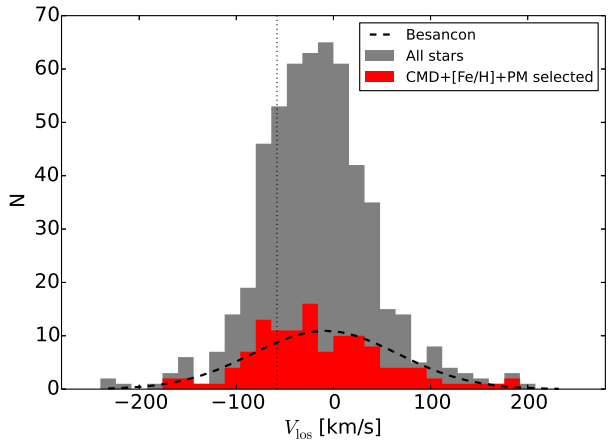}
\caption{Line-of-sight velocity histograms of the sample stars with $S/N>10$ in the 10 DEIMOS fields, The gray and red histograms are for all stars and candidate stream stars, respectively.
  The vertical dotted line corresponds to the velocity of the central cluster ($-58.1$ km s$^{-1}$). The dashed line represents the line-of-sigh velocity distribution at the direction of FD4 for the field MW stars based on the Besan\c{c}on model.  }
\label{fig:vrdistall}
\end{figure}

Figure \ref{fig:vrdistall} shows line-of-sight velocity distributions for
the stars in all of the DEIMOS fields.
The $V_{\rm los}$ distribution for the all sample stars
(S/N$>10$, $V_{\rm los}$ error $<10$ km s$^{-1}$; the gray histogram) is
 dominated by the field MW stars and peaked at $\sim-20$ km s$^{-1}$. 
 In the distribution for the sample stars satisfying the criteria 1-3 (the red
 histogram),
 the velocity peak dominated by
the field stars is suppressed but still remains to be significant.

For comparison, the expected distribution for the field MW stars satisfying the
same CMD and [Fe/H] criteria (criteria 1 and 2)  
is shown in the dashed line in Figure \ref{fig:vrdistall}.
The distribution of the field MW stars is peaked at $-9$ km s$^{-1}$
with dispersion of $77$ km s$^{-1}$.
The $V_{\rm los}$ distribution likely representing the stream stars
is peaked at $\sim -65$ km s$^{-1}$, which lies within the
velocity distribution of the field MW stars, preventing a clean
separation of the member stars. 
If we apply an additional $V_{\rm los}$ cut, $-100<V_{\rm los}<-20$ km s$^{-1}$,
to minimize contamination from the field MW stars and to be $\pm \sim 40$
km s$^{-1}$ of the $V_{\rm los}$ of the Pal 5 cluster ($-58$ km s$^{-1}$),
54 stars remain.
 Positions, magnitudes, line-of-sight velocities, metallicites
  and the status of the $V_{\rm los}$ cut for stars satisfying
  the criteria 1-3 are given in Table \ref{tab:selectedstars}.

The left panel of Figure \ref{fig:cmd_candidate} shows
a CMD of the target stars. The candidate stream stars, which satisfy
the criteria 1-3 and have $-100<V_{\rm los}<-20$ km s$^{-1}$, are
marked as red circles. For a comparison, the right panel
shows an expected distribution of the
field MW stars around one of the DEIMOS field, FD4,
obtained by the Besan\c{c}on model \citep{robin03}.

The major contaminants
to the candidate stream stars are the
MW field thick disk and halo stars. We estimate the expected number of
contaminating stars using the Besan\c{c}on model by applying the same
cuts as the sample stars and scaling the number of stars according
to the field-of-view of the DEIMOS mask (16.7\arcmin$\times$5\arcmin).
Taking into account the fact that
our spectroscopic
sample with the adopted quality limit
(S/N$>10$ and $V_{\rm los}$ error $<10$ km s$^{-1}$) amount to 18-44 \%
of the SDSS photometric sample in the magnitude range $r=14-21$ in the
ten DEIMOS fields, the expected number of contaminants is 35 or 40,
depending on whether or not the proper motion criteria 3 is 
applied. This implies that more than half of the 54 candidate stream
stars could be contaminants from the field MW stars, while 
 a certain fraction of
them could actually belong to the stream.

\begin{deluxetable*}{cccccccc}
  \tablecaption{Stars selected by the CMD, [Fe/H] and proper motion criteria.\label{tab:selectedstars}}
  \tablewidth{0pt}
  \tablehead{\colhead{Field} & \colhead{Star} & \colhead{RA} & \colhead{DEC} & \colhead{$g_{0}$\tablenotemark{a}} & \colhead{$V_{\rm los}$}  & \colhead{[Fe/H]} & \colhead{$V_{\rm los}$ cut\tablenotemark{b}} \\
  \colhead{} & \colhead{} & \colhead{(deg)} & \colhead{(deg)} & \colhead{(mag)} & \colhead{(km s$^{-1}$)} & \colhead{(dex)} & \colhead{}}
\startdata
FD1 & 002 & 227.30743 & $  -2.0780$ & 20.23 & $   86.0\pm 4.6$ & $  -1.08$ & n \\
FD1 & 006 & 227.25285 & $  -2.0564$ & 17.82 & $   53.6\pm 3.0$ & $  -1.08$ & n \\
FD1 & 007 & 227.29528 & $  -2.0668$ & 17.38 & $  -72.7\pm 1.7$ & $  -1.09$ & y \\
FD1 & 008 & 227.23506 & $  -2.0649$ & 20.69 & $  -68.7\pm 9.2$ & $  -1.61$ & y \\
FD1 & 011 & 227.36408 & $  -2.0007$ & 19.28 & $   16.6\pm 3.3$ & $  -1.58$ & n 
\enddata
\tablecomments{Table \ref{tab:selectedstars} is published in its entirety in the 
electronic edition of the {\it Astrophysical Journal}.  A portion is 
shown here for guidance regarding its form and content.}
\tablenotetext{a}{Extinction-corrected SDSS $g$-band magnitude.}
\tablenotetext{b}{'y' for an object with $V_{\rm los}$ in a range $-100<V_{\rm los}<-20$ km s$^{-1}$, 'n' otherwise.}  
\end{deluxetable*}

\subsection{Individual fields}

\begin{figure}
\begin{center}
\includegraphics[width=8.5cm]{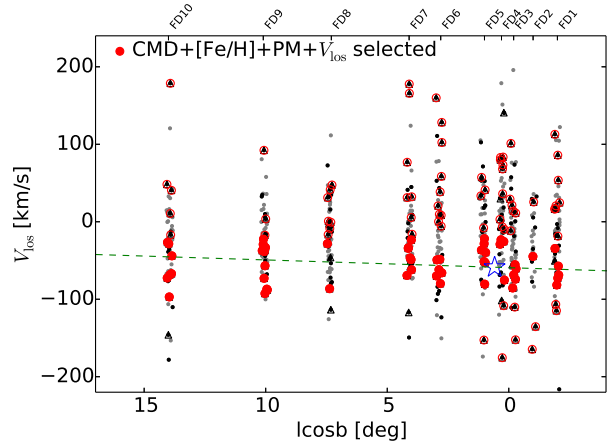}
\caption{The line-of-sight velocities 
  of the sample stars as a function of $l\cos b$. All stars with S/N$>10$
  are shown in gray dots. 
Among them, stars satisfying our candidate selection criteria are 
shown in different symbols: (1) the CMD criterion 
 (black dots), (2) the CMD and [Fe/H] criteria (open 
triangles), (3) the CMD, [Fe/H] and proper motion criteria 
(red circles), among which objects having $-100<V_{\rm los}<-20$ km s$^{-1}$
are shown in filled red circles.
The $V_{\rm los}$ of the central Pal 5 cluster is marked
by a star symbol. The dashed line indicates the $V_{\rm los}$ gradient
suggested in the previous studies \citep{odenkirchen02,kuzma15}.}
\label{fig:rv_lcosb}
\end{center}
\end{figure}

Figure \ref{fig:rv_lcosb} shows the 
line-of-sight velocities of the sample stars 
as a function of $l\cos b$.  
The gray dots indicate all stars and  
the black dots indicate stars that satisfy the color-magnitude 
criterion 1. The open triangles
are for objects that fulfill the 
criteria 1 and 2 but not the criterion 3. Finally, the red   
circles represent objects that satisfy all of the three criteria.
Among them, the objects with $-100<V_{\rm los}<-20$ km s$^{-1}$
are shown in filled circles (in the following, we refer these stars
as {\it "candidate stars''}). The $V_{\rm los}$ distribution of the
stars selected by the
criteria 1-3 have a wide range of velocities, again suggesting
a heavy contamination of field MW star.

The $V_{\rm los}$ distributions of stars that fulfill the
criteria 1-3 in each of the DEIMOS fields are shown in
Figure \ref{fig:vrdist}. These are compared with
expected velocity distributions of the contaminating
field stars from the Besan\c{c}on model. 
By applying similar selection criteria as those
adopted for the target stars, the $V_{\rm los}$ distribution for the
thick disk stars (blue dotted line) is peaked at $\sim 0$ km s$^{-1}$.
The distributions of the
halo stars are peaked at negative $V_{\rm los}$ values and thus
remains as major contaminants after applying the $V_{\rm los}$ cut. 
The halo star contamination is especially important at FD8-FD10,
in which velocity distribution peaks of the halo stars
lie between $-60$ and $-30$ km s$^{-1}$.

After applying the $V_{\rm los}$ cut, although the contaminants
may still exist, 
the candidate stars have larger negative values of $V_{\rm los}$
than the Pal 5 cluster on the leading tail. The more distant field (FD1) has
candidate stars with larger negative velocities, which would be 
consistent with the $V_{\rm los}$ gradient along the leading tail.
The candidate stars on the trailing tail, on the other hand
have $V_{\rm los}$ with smaller negative velocity
than the Pal 5 cluster and show a sign of
gradient toward more distant
part of the stream, as far as the range $-3<l\cos b<5$\arcdeg is concerned.
For more outer part of the stream (FD8-FD10),
the signature of the gradient is ambiguous and
some of the candidate stars have $V_{\rm los}<-60$ km s$^{-1}$.

\subsection{The $V_{\rm los}$ range for the stream}
\label{sec:gradient}

As mentioned in the previous sections, it is not possible 
to individually separate genuine stream stars from the field MW 
stars because of the similarity in both $V_{\rm los}$ and [Fe/H]
between the stream and the field stars.
In this subsection, 
we make a statistical inference of likely $V_{\rm los}$ values of the 
stream at each DEIMOS field, instead of applying the sharp [Fe/H] and $V_{\rm los}$ cuts. We construct a model for line-of-sight velocities
and metallicities of the CMD-selected stars (the criterion 1 in Section
\ref{sec:selection} in the each DEIMOS field 
as a combination of actual stream stars and 
foreground/background MW stars by the method similar 
to that applied in \citet{walker11}.

We first assume that, in each of the DEIMOS fields, the 
stream stars are characterized by 
a mean line-of-sight velocity ($\bar{V}$), 
its intrinsic dispersion ($\Sigma_{V}$) and a fraction $f_{{\rm st}}$ 
of the stream stars among the CMD-selected stars. Metallicities
of the stream stars are assumed to be identical to that of the 
central cluster estimated in Section \ref{sec:pal5cen} within the uncertainty.
A likelihood of obtaining a line-of-sight velocity $V_{i}$ 
and a metallicity $M_{i}$ for an $i$'th sample star given the model parameters 
$\Theta=\{\bar{V}, \Sigma_{V}, f_{{\rm st}}\}$, is expressed as,

\begin{eqnarray}
L_{i}\left( V_i, M_i | \Theta\right)=f_{{\rm st}}P_{{\rm st}}(V_i,M_i)+\nonumber\\
(1-f_{\rm st})P_{\rm MW}(V_i,M_i)
\label{eq:likeik}
\end{eqnarray}

$P_{\rm st}(V_i,M_i)$ is a probability of a stream star to have measured 
line-of-sight velocity and metallicity, $V_i$ and $M_i$, respectively in 
each field. 

These probabilities are given by: 

\begin{eqnarray}
  P_{\rm st}(V_i,M_i)&=&P_{V, {\rm st}}(V_i) P_{M, {\rm st}}(M_i)
\end{eqnarray}

where

\begin{eqnarray}
P_{V, {\rm st}}(V_i)&=&\frac{1}{\sqrt{2\pi(\Sigma_{V}^2+\epsilon_{V,i}^2)}}\exp\left[-\frac{1}{2}\frac{(V_i-\bar{V})^2}{\Sigma_{V}^2+\epsilon_{V,i}^2}\right]\nonumber\\ 
P_{M, {\rm st}}(M_i)&=&\frac{1}{\sqrt{2\pi\epsilon_{M,i}^2}}\exp\left[-\frac{1}{2}\frac{(M_i-\bar{M})^2}{\epsilon_{M,i}^2}\right]\nonumber
\end{eqnarray}
 
where $\epsilon_{V,i}$ and $\epsilon_{M,i}$ are the uncertainties in 
velocity and metallicity measurements, respectively, and $\bar{M}$ is the
metallicity of the Pal 5 cluster.

The corresponding probability for the foreground/background MW stars 
is expressed as

\begin{eqnarray}
P_{\rm MW}(V_{i}, M_{i})=(1-f_{2}-f_{3})P_{1}(V_i, M_i)+\nonumber\\
f_{2}P_{2}(V_i, M_i)+f_{3}P_{3}(V_i, M_i)
\end{eqnarray}

where 
\begin{eqnarray}
P_{j}(V_i,M_i)&=&P_{V,j}(V_i)P_{M,j}(M_i) \\
P_{V,j}(V_i)&=&\frac{1}{\sqrt{2\pi(\sigma_{V,j}^2+\epsilon_{V,i}^2)}}\exp\left(-\frac{1}{2}\frac{(V_i-\bar{V_{j}})^2}{(\sigma_{V,j}^2+\epsilon_{V,i}^2)}\right) \nonumber\\
P_{M,j}(M_i)&=&\frac{1}{\sqrt{2\pi(\sigma_{M,j}^2+\epsilon_{M,i}^2)}}\exp\left(-\frac{1}{2}\frac{(M_i-\bar{M_{j}})^2}{(\sigma_{M,j}^2+\epsilon_{M,i}^2)}\right). \nonumber
\end{eqnarray}

In the above expressions, $j=1,2,$ and $3$, correspond to the thin disk, thick disk and halo components, respectively.
The mean line-of-sight velocity ($\bar{V_{j}}$) and 
its dispersion ($\sigma_{V,j}$) for each Galactic component 
as well as  $f_{2}$ and $f_{3}$ 
are evaluated in advance based on the Besan\c{c}on model. In this
step, the mean metallicity ($\bar{M_j}$) and dispersions ($\sigma_{M,j}$)
are fixed to the values 
$(\bar{M_{j}},\sigma_{M,j})=(-0.1,0.2), (-0.8,0.3), (-1.8,0.5)$ for the 
thin disk, thick disk and halo, respectively. 

A likelihood for each field $k$ is obtained by multiplying the 
expressions in (\ref{eq:likeik}) for the number of stars ($N_{k}$), 
\begin{equation}
L=\prod^{N_k}_{i=1}L_{i}
\end{equation}

\begin{figure}
\begin{center}
\includegraphics[width=8.5cm]{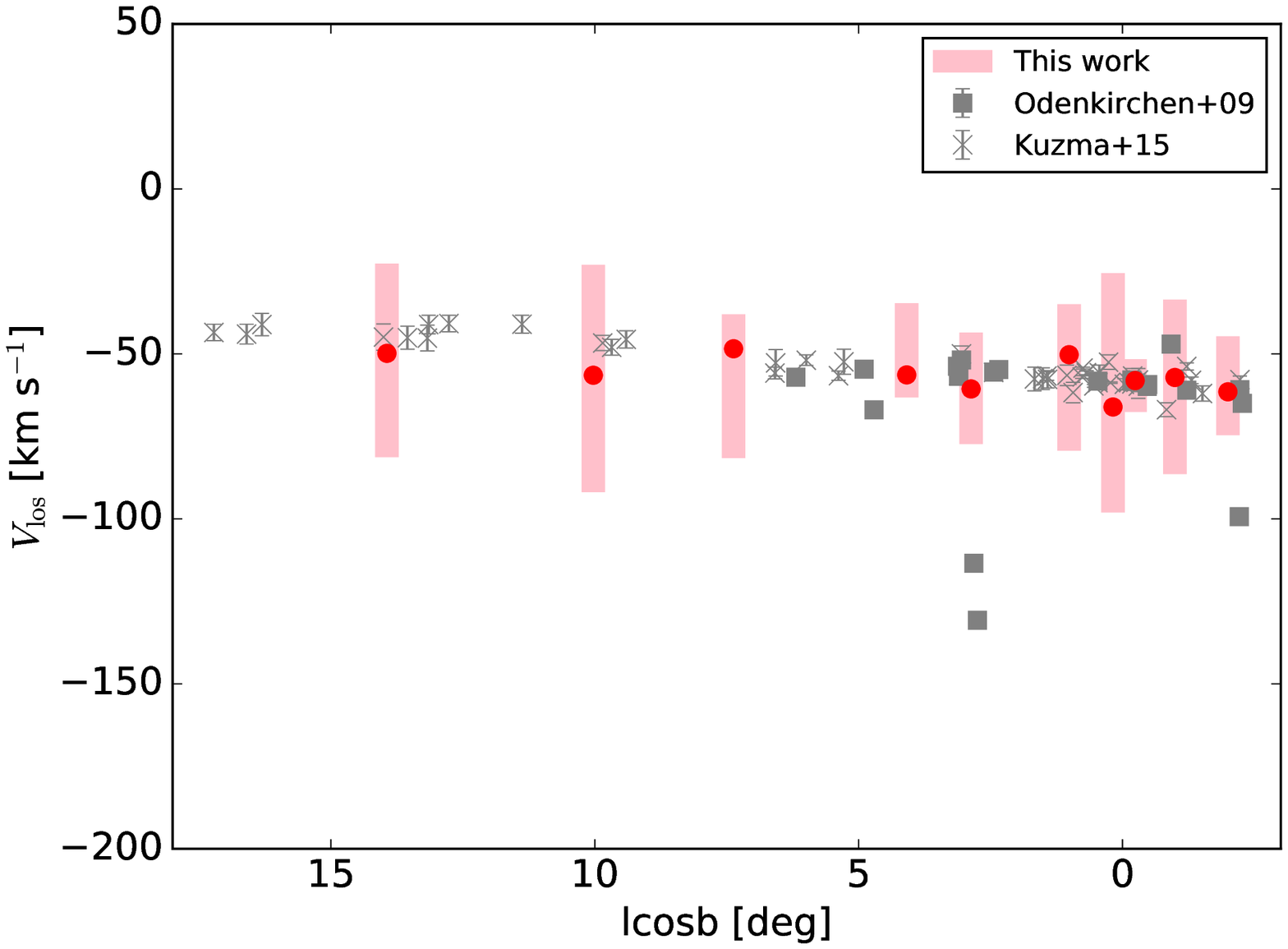}
\caption{The likely $V_{\rm los}$ ranges for the Pal 5 stream in
  each DEIMOS field. Red points and pink rectangles correspond to the value of the
  50th percentile and the ranges between
  the 16th and 84th percentiles, respectively, for the posterior distributions of the
  $V_{\rm los}$ obtained in this work.
The $V_{\rm los}$ of candidate member stars
  identified in \citet{odenkirchen09} and \citet{kuzma15} are also indicated with squares and crosses, respectively.
}
\label{fig:lcosb_vlos_linfit}
\end{center}
\end{figure}

Posterior distributions for 
the parameter-set $\Theta$ given the data $\{V_i,M_i\}_{i=1}^{N_k}$ is 
expressed as, 

\begin{equation}
p(\Theta | \{V_i,M_i\}_{i=1}^{N})\propto L(\{V_i,M_i\}_{i=1}^{N_k}|\Theta) I(\Theta)
\end{equation}

\noindent
where $I(\Theta)$ is a prior probability distribution for the parameter 
set. For the $I(\Theta)$ , uniform distributions in the ranges 
$-100< \bar{V_k} < -20$ km s$^{-1}$, $0<\Sigma_{V, k}<5$ km s$^{-1}$
  and $0.0<f_{{\rm st}, k}<0.5$ are adopted, where the former for $\bar{V_k}$
  is symmetric around the mean velocity of the Pal 5 cluster
  ($\sim -58$ km s$^{-1}$) and the latter for $f_{{\rm st}, k}$ is
aimed to cover the likely range of the stream fraction.
The Markov-Chain Monte Carlo code, {\it emcee}, \citep{foreman-mackey13}
is used to sample the 
posterior probability distribution.
The resulting posterior distributions of $\bar{V_k}$,
marginalized over the $\Sigma_{V, k}$ and $f_{{\rm st}, k}$, do not
always have a well-defined single peak but end up with multiple
peaks. The point of 50th percentile and the ranges between the 16th and 84th percentiles in the resulting distributions of $V_{\rm los}$ are
shown by red circles and pink rectangles, respectively, in Figure
\ref{fig:lcosb_vlos_linfit}.

In the DEIMOS fields at $l\cos b<5$\arcdeg, the likely $V_{\rm los}$
values for the stream well overlap with the member stars identified
in the previous works \citep{odenkirchen09,kuzma15}. 
In the outer region ($l\cos b>5$\arcdeg), FD8 and FD10,
the likely $V_{\rm los}$ values for the
stream also overlap with the member stars identified by \citet{kuzma15}
while allowing for lower $V_{\rm los}$ than that expected from
an extrapolation of the $l\cos b-V_{\rm los}$ relation found in $l\cos b<5$\arcdeg.

\begin{figure}
\includegraphics[width=8.5cm]{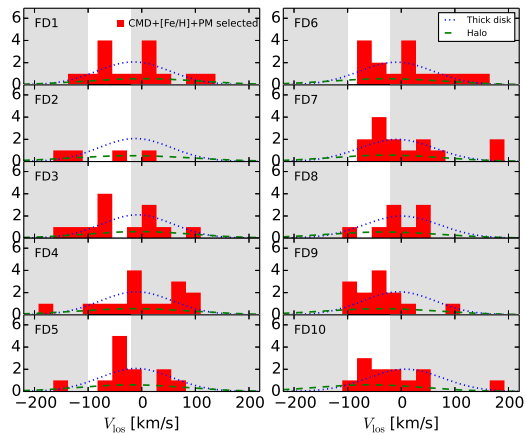}
\caption{Line-of-sight velocity histograms (red) for stars selected by the criteria (1)-(3) in Section \ref{sec:selection}. The white region corresponds
  to the $V_{\rm los}$ limit of $-100<V_{\rm los}<-20$ km s$^{-1}$.
  The dotted and dashed lines correspond to
  the velocity distributions (arbitrary vertical scale)
  for the simulated field thick disk and halo stars,
  respectively, obtained from the Besan\c{c}on model.}
\label{fig:vrdist}
\end{figure}

We note that the prior assumption about $\bar{V_k}$ is crucial in
obtaining the results mentioned above. In order to
check the robustness of the result, we repeat the 
same analyses by relaxing the prior assumptions on the
possible value of $\bar{V_k}$ to $-160< \bar{V_k} < 0$ km s$^{-1}$.
The resulting distributions show another peak
within a range from $\sim 0$ km s$^{-1}$ at FD8 to $-30$ km s$^{-1}$
at FD10. These peaks roughly coincide with the peak in the velocity distribution
of the thick disk and halo stars (Figure \ref{fig:vrdist}). Since the 
$l\cos b-V_{\rm los}$ trend of that peak is in opposite sense to
that expected
for the Pal 5 stream, it is less likely to be associated with the stream.
We also tests a narrower prior of $-100<\bar{V_k}<-30$ km s$^{-1}$
  to reduce
contamination from the field thick disk stars. This results in more
negative ranges for the posterior $\bar{V_k}$ distribution in the
outer part of the stream, which highlight the importance of
the adopted prior on the resulting inference of
$\bar{V_k}$. 

The high fraction of field MW stars relative to the stream stars
in our sample could be understood as an actual
deficiency in relatively bright (higher-mass)
stars in the stream as reported by \citet{koch04} together with
the shallow depth in the present spectroscopic observation
(up to $r\sim 20.5$). This situation should be improved with deeper
spectroscopic observations that can reach
below the main-sequence turn-off, which would significantly increase 
a fraction of the stream stars relative to the field MW stars.

\section{Discussion}
\label{sec:discussion}

\subsection{Comparison with previous studies}

Line-of-sight velocities based on high-resolution
spectroscopy have previously been reported by \citet{odenkirchen09} covering
an angular extent of $\sim 8.5$\arcdeg of the streams.
Candidate member stars in the streams are selected based on
the Mg b triplet feature around $5180$ {\AA}, which is an indicator
of a stellar surface gravity. The line-of-sight velocity distribution
of the target stars that are identified as a giant is
found to be mostly concentrated in a range $-70$ to $-40$ km s$^{-1}$ and
peaks at $-58$ km s$^{-1}$.
The peak velocity is consistent with that of the central cluster,  
suggesting that these giants are indeed plausible candidate members of the
stream.
Based on these identified members, 
they reported a $V_{\rm los}$
gradient of  $1.0\pm 0.4$ km s$^{-1}$ deg$^{-1}$ as a function of
the Galactic longitude.  

The small but non-negligible
velocity gradient is supported by 
\citet{kuzma15}, who have identified candidate member stars over a much
larger angular extent.
\citet{kuzma15} make use of the \ion{Ca}{2} triplet absorption lines 
as a proxy of metallicity ([Fe/H]) to identify probable member stars. 
After applying their membership criteria, including the \ion{Ca}{2} equivalent
widths, dwarf/giant 
separation, the CMD and the line-of-sight velocities (assumed to
be in the range from $-70$ to $-35$ 
km s$^{-1}$), 47 candidate stars have been identified over 
an angular extent $\sim 20$\arcdeg along the stream. 
The best-fit linear gradient for these stars
is 1.0$\pm$0.1 km s$^{-1}$ deg$^{-1}$ , which
is in agreement with \citet{odenkirchen09}.

As mentioned in the previous section, the likely $V_{\rm los}$ range
at each stream locations in this work taking into account
the contamination of the field stars well overlap with the
$V_{\rm los}$ of the member stars identified in \citet{odenkirchen09}
and \citet{kuzma15} for $l\cos b<5.0$\arcdeg. For this angular extent,
the present result is also compatible with the presence of
a gradient suggested by their studies.

In the more outer part of the trailing stream
(FD8-FD10), the $V_{\rm los}$ ranges also overlap
with those of the identified member stars in \citet{kuzma15},
while they allow 
lower $V_{\rm los}$ values for the stream such that $V_{\rm los}<-50$ km s$^{-1}$. 
This result would be explained by either 
(1) significant contamination of the field stars 
or (2) deviation in $V_{\rm los}$ of the stream stars from that
of the Pal 5 cluster orbit.

In the former case, the small 
field-of-view of a single DEIMOS exposure 
might have hampered detection of 
the genuine member stars. As discussed in Section \ref{sec:selection},
contamination of the field stars of up to $\sim 80$ \% is expected
according to the Besan\c{c}on model and thus it is not
surprising that the stars that fulfill the adopted criteria
in each DEIMOS field are dominated by the contaminants.

For the latter case, some stars once belonging
to the cold part of the stream have gained larger
velocity dispersions, resulted from e.g. 
dynamical interactions with dark matter subhalos, which are presumably
ubiquitous according to the currently standard $\Lambda CDM$ cosmology
\citep[e.g.][]{yoon11,bonaca14,ngan16}. 
\citet{yoon11} suggests that the interaction with the dark
matter subhalos results in change in $V_{\rm los}$ and velocity
dispersion along a cold stellar stream,  
depending on the masses of the interacting dark matter
subhalos and frequency of the encounters.
As an alternative possibility, \citet{pearson15} suggests that the
morphology and velocity of a stream like Pal 5 would be significantly
modified, if the potential of the MW's dark matter halo
was triaxial like the one suggested by \citet{law10}.  
It is not clear, however, that whether the possible 
perturbations to the $V_{\rm los}$ is compatible with
the observed narrowness in the morphology of the stream. 
Indeed, \citet{kuzma15} suggest that the stream is kinematically
  cold even in the outer region, which does not support the significant
  heating.

In order to evaluate presence or absence of significant
perturbations on the stream's line-of-sight velocities,
analyses of a larger samples without a
strong bias in $V_{\rm los}$ itself are required. 
Identification of the stream member stars, therefore, using chemical abundances
(metallicities) as well as proper motions would be crucial to evaluate the possible
dynamical heating experienced by Pal 5. 

\subsection{Implications for the Galactic potential}

In this section, we use the possible $V_{\rm los}$ ranges along the
Pal 5 stream obtained in Section \ref{sec:gradient} to investigate
models of the MW's gravitational potential that are compatible with the data.  

We assume that the Galactic potential is described as a 
sum of three components, namely, the bulge, disk and dark halo.  
We adopt the Hernquist model \citep{hernquist90}, axis-symmetric 
Miyamoto-Nagai potential \citep{miyamoto75} and the Navarro-Frenk-White
(NFW) profile \citep{navarro97} for the bulge, disk and 
dark halo components, respectively, as adopted in \citet{kupper15}.

\begin{eqnarray}
\Psi_{\rm bulge}(r)&=&-\frac{GM_{\rm bulge}}{r+a}\nonumber\\
r&=&\sqrt{\mathstrut x^2+y^2+z^2}\nonumber\\
\Psi_{\rm disk}(R,z)&=&-\frac{GM_{\rm disk}}{\sqrt{\mathstrut R^2+\left(b+\sqrt{\mathstrut z^2+c^2}\right)^2}}\nonumber \\
R&=&\sqrt{\mathstrut x^2+y^2}\nonumber \\
\Psi_{\rm halo}(r)&=&-\frac{GM_{\rm halo}}{r}\ln \left(1+\frac{r}{r_{\rm halo}}\right) \nonumber \\
r&=&\sqrt{\mathstrut R^2+\frac{z^2}{q_{\rm halo}^2}}
\end{eqnarray}  
where $G$ is a gravitational constant, with $M_{\rm bulge}=3.4\times 10^{10} M_{\odot}$, $a=0.7$ kpc, $M_{\rm disk}=10^{11}M_{\odot}$, $b=6.5$ kpc and $c=0.26$ kpc.

Position and kinematics of the Pal 5 cluster are adopted mainly
from the values derived in recent literature. 
The distance to the Pal 5 cluster $d_{\rm pal5}=23.5$ kpc is adopted from the
$V$ band distance modulus based on isochrone fitting to the HST
photometric data \citet{dotter11}. 
The proper motion of Pal 5 has recently been reported to be    
$(\mu_{\alpha}, \mu_{\delta})=(-2.296\pm 0.186, -2.257\pm 0.181)$ mas yr$^{-1}$
based on combined SDSS and Large Binocular Telescope/Large Binocular Camera
data, which span a baseline of 15 years \citep{fritz15}.
To make a comparison with \citet{kupper15}, we first adopt the proper
motions obtained in their work, $(\mu_{\alpha}\cos (\delta), \mu_{\delta})=(-2.40, -2.38)$, then update these values to the latest estimates of
\citet{fritz15}. 

For the solar-tangential velocity and the distance from the
Galactic center, we adopt $V_{\rm tan}=V_{\rm LSR}+V_{\odot}=255.2$ km s$^{-1}$
and $R_{\odot}=8.34$, respectively from \citet{reid14}.  
The radial and vertical components of the solar motion is adopted to be 
$U_{\odot}=11.1$ and $W_{\odot}=7.3$ km s$^{-1}$, which were derived by
\citet{schonrich10}.

We investigate orbits in the potential described above for various
dark halo parameters while fixing the model parameters for the bulge
and disk components as well as the solar position and kinematics. Specifically,
the dark halo parameters obtained by \citet{kupper15} using
the previously available $V_{\rm los}$ data of \citet{odenkirchen09},
with their "Overdensity + Radial Velocities'' method ($M_{\rm halo}=1.75^{+0.76}_{-0.66}$ [10$^{12} M_{\odot}$], $r_{\rm halo}=41.8^{+14.5}_{-11.0}$ [kpc], and $q_{\rm halo}=0.84^{+0.27}_{-0.16}$) are considered as a standard set of values.
We calculate the orbits by varying one of the $M_{\rm halo}$, $r_{\rm halo}$ and $q_{\rm halo}$ from the standard set, while the other two are fixed.

\begin{figure}
\begin{center}
\includegraphics[width=8.5cm]{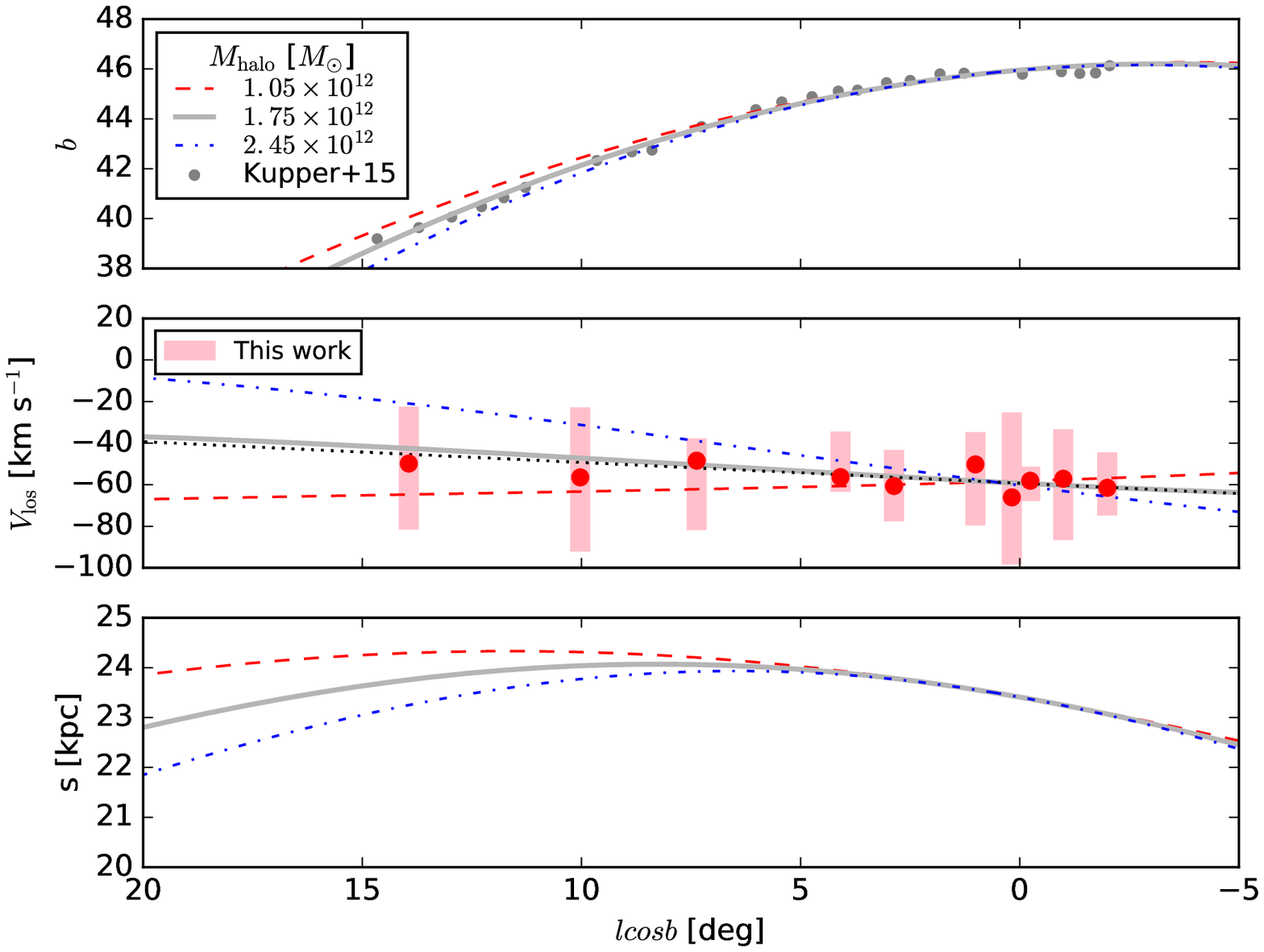}
\caption{The model orbits calculated based on the parameters estimated by \citet{kupper15} (the thick gray line; $M_{\rm halo}=1.75\times 10^{12} M_{\odot}$) and for different $M_{h}$ (the red dashed line for $1.05\times 10^{12} M_{\odot}$ and blue dash-dotted line for $2.45\times 10^{12} M_{\odot}$) are compared with the observed positions \citep{kupper15} and the $V_{\rm los}$ range obtained in this work. The top, middle and bottom panels show Galactic latitude ($b$), line-of-sight velocities ($V_{\rm los}$) and distances ($s$) as functions of Galactic longitude ($l cosb$). The black dotted line indicates the linear fit to the $V_{\rm los}$ data obtained in \citet{kuzma15}. The estimated range of $V_{\rm los}$ at each location along the stream are shown with pink rectangles.}
\label{fig:modelstream_mh}
\end{center}
\end{figure}

Figure \ref{fig:modelstream_mh} shows the resulting orbits for different
$M_{\rm halo}$. The top, middle and the bottom panels show the Galactic latitude ($b$), $V_{\rm los}$, and distance from the Sun ($s$), respectively,
as a function of $l\cos b$. As can be seen from the middle panel,
the models for  $M_{\rm halo}=1.75$ (thick gray line) and
$1.05\times 10^{12} M_{\odot}$ (red dashed line),
which correspond to a circular velocity at the solar radius
$V_{c}(R_{\odot})=231$ and $218$ km s$^{-1}$, respectively, lie within the likely $V_{\rm los}$ ranges obtained in this work
(pink rectangles). On the other hand the higher $M_{\rm halo}$ value of
$2.45\times 10^{12}M_{\odot}$ (blue dash-dotted line), which corresponds to
$V_{c}(R_{\odot})=242$ km s$^{-1}$, is incompatible with the data
in the outer part of the stream.

\begin{figure}
\begin{center}
\includegraphics[width=8.5cm]{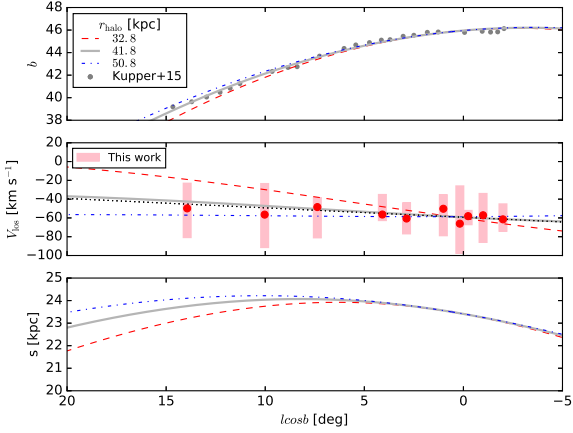}
\caption{Same as in Figure \ref{fig:modelstream_mh} but for orbits with different $r_{\rm halo}$ (the thick gray line for $41.8$ kpc, the red dashed line for $32.8$ kpc, and the blue dash-dotted line for $50.8$ kpc). }
\label{fig:modelstream_rh}
\end{center}
\end{figure}

Figure \ref{fig:modelstream_rh} shows the orbits calculated
varying $r_{\rm halo}$ by $\pm 9$ kpc from the standard value,
while the other parameters are fixed. 
The two models for the larger $r_{\rm halo}$, which
correspond to smaller $V_{c}(R_{\odot})$ values ($\lesssim 231$),
are compatible with the estimated $V_{\rm los}$ ranges. 
On the other hand, the smaller $r_{\rm halo}$
value predicts much higher $V_{\rm los}$ than those suggested
in the present analysis.

\begin{figure}
\begin{center}
  \includegraphics[width=8.5cm]{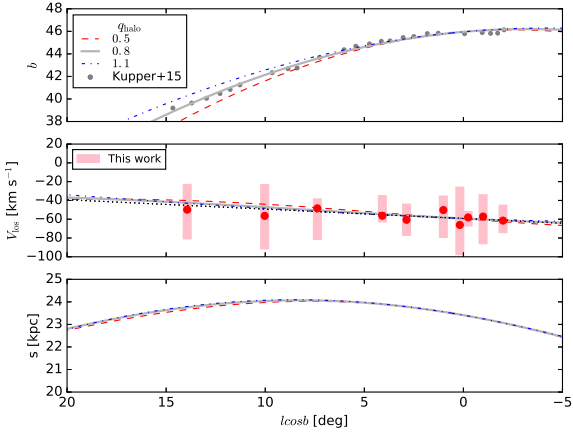}
\caption{Same as in Figure \ref{fig:modelstream_mh} but for orbits with different $q_{\rm halo}$ (the thick gray line for 0.8, the red dashed line for 0.5, and the blue dash-dotted line for 1.1).}
\label{fig:modelstream_qh}
\end{center}
\end{figure}

For the models considered in this work, the changes in $q_{\rm halo}$
by $\pm 0.3$ from the standard value of 0.8
have only a small effect on the $V_{\rm los}$ when other parameters
are kept fixed, as shown in Figure \ref{fig:modelstream_qh}.
Larger changes in $q_{\rm halo}$ would result in significant
discrepancy between the predicted and
the observed projected locations of the stream as can be seen
in the top panel.

\begin{figure}
\begin{center}
  \includegraphics[width=8.5cm]{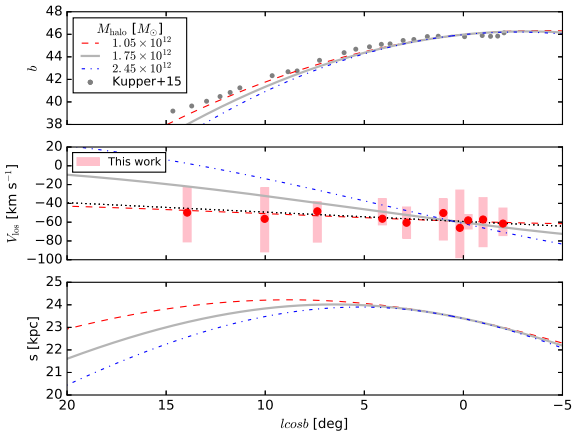}
\caption{Same as in Figure \ref{fig:modelstream_mh} but for orbits with updated proper motion values from \citet{fritz15}.}
\label{fig:modelstream_PM}
\end{center}
\end{figure}

Finally, Figure \ref{fig:modelstream_PM} shows the same models as
in Figure \ref{fig:modelstream_mh} but with the updated values of
proper motion of Pal 5 cluster's obtained
by \citet{fritz15}. The
absolute values of their best proper motion estimates are slightly
smaller by 0.10 and $-0.12$ mas yr$^{-1}$ for $\mu_{\alpha}\cos\delta$
and $\mu_{\delta}$,
respectively, than those adopted above although these estimates are
consistent within the uncertainty.
As shown in the middle panel,
  the smallest $M_{\rm halo}$ model, which corresponds to
$V_{c}(R_{\odot})=218$
km s$^{-1}$, is compatible with the $V_{\rm los}$
range, while the $M_{\rm halo}=1.75\times 10^{12} M_{\odot}$ model is
only marginally consistent with the data
in the outer-most field on the stream. The $V_{\rm los}$ estimates
combined with the updated proper motion data, therefore, are
better reproduced with the models with $V_{c}(R_{\odot})<231$
km s$^{-1}$, for the given choice of other MW parameters.
However, as discussed in \citet{fritz15}, the conclusion on the
model could also depend on the adopted distance to Pal 5 cluster. 
Proper motion and distance measurements with improved precision are
essential to conclude on whether or not the models with larger $M_{\rm halo}$
are ruled out. 

Nevertheless, the preference for the relatively small $V_{c}(R_{\odot})$ is
in line with the recent distance estimates by
\citet{ibata15} along the Pal 5 stream. Their analyses
report continuous increase in the distances
to the stream from the leading to the trailing tails
up to $\sim 10$\arcdeg from the Pal 5 cluster. 
This trend is reproduced with the models with smaller $V_{c}(R_{\odot})$
(either smaller $M_{\rm halo}$ or larger $r_{\rm halo}$),
as shown in the bottom panels in
Figure \ref{fig:modelstream_mh}, \ref{fig:modelstream_rh} and \ref{fig:modelstream_PM}.
On the other hand, the models with the larger $V_{c}(R_{\odot})$
values (either $M_{\rm halo}=2.45\times 10^{12}M_{\odot}$ or
$r_{\rm halo}=32.8$ kpc)
predict a turn around in the distance trend at a point
closer to the Pal 5 cluster.

To summarize, for the given values for the bulge and disk
parameters as well as the solar position and kinematics,  
the models for the dark matter halo with
$V_{c}(R_{\odot})\lesssim 231$ km s$^{-1}$ are consistent with
the estimated range for the possible $V_{\rm los}$ values for the Pal 5 stream.
This result, however, depends on various factors such as
the prior assumption about the $V_{\rm los}$ for the
stream and the adopted value for the proper motion of the Pal 5
cluster. The uncertainty in $V_{\rm los}$ primarily comes from
the fact that the likely $V_{\rm los}$ and the 
[Fe/H] of the stream members are similar to those for the
field MW stars, and thus, with a small sample size, it is
difficult to separate the two contributions. 
Planned and ongoing wide-field and deeper 
spectroscopic surveys with multi-object capabilities such as LAMOST, WHT/WEAVE,
4MOST, Subaru/PFS would provide
chemical compositions and $V_{\rm los}$ together for a large number of
stars, which offer an excellent opportunity to better understand
the nature of streams and their constraints on the MW's dark matter halo.

\section{Conclusions}

We carried out multi-object, medium-resolution spectroscopy
along the central cluster and the tidal stream of Pal 5, which
enables $V_{\rm los}$ and [Fe/H] measurements over an
angular extent of $\sim 17$\arcdeg along the stream.  
By adopting the spectral fitting technique to the observed
spectra, together with the available SDSS
photometry, we derived stellar parameters and [Fe/H] of
the sample stars.
The estimated $V_{\rm los}$ 
and [Fe/H] for the central cluster are $-58.1\pm 0.7$ km s$^{-1}$ and [Fe/H]$=-1.35\pm 0.06$
dex, respectively, 
which are in good agreement with that derived in previous
high-resolution spectroscopic studies.
This value of [Fe/H] for the central cluster 
is used first to identify  candidate member stars
and second to estimate
the range in possible $V_{\rm los}$ of the stream at each projected location
along the stream.

The resulting ranges of $V_{\rm los}$ depend on the prior assumptions
about $V_{\rm los}$ of the stream because of the
overlap in both the expected $V_{\rm los}$ and [Fe/H]
of the stream with that of
the field MW thick disk/halo stars. By assuming the stream $V_{\rm los}$
to be $-100<V_{\rm los}<-20$ km s$^{-1}$, the inferred 
$V_{\rm los}$ range is consistent with previous studies,
while it does not exclude the possibility of lower $V_{\rm los}$
at the outer part of the stream. These analyses
highlight the importance of more definitely
identifying the member stars by using e.g. proper motion and 
detailed chemical information, in order to obtain unbiased
estimates of $V_{\rm los}$ for the entire part of the Pal 5 system.
Wider and deeper spectroscopic surveys that provide
both $V_{\rm los}$ and chemical compositions for a large
number of individual stars would provide important
data to better understand the kinematics of the stream, which
might eventually put stronger constraint on the nature of
the MW's dark matter halo.

\acknowledgments
The authors are grateful to the referee for his/her constructive
comments which helped to improve this paper. 
We thank T. Hattori and other staff members in Subaru Telescope
for their various technical supports in preparing
and carrying out our observations.  
We are also grateful to support astronomers and other staff members of
Keck telescope for their generous supports for the observation
and data analysis. M.N.I. thank E. Kirby and J. Cohen for their
helpful advice on spectroscopic analyses and A. More for
helpful discussions during the early
stage of this work. 
M.N.I. acknowledge support through
JSPS KAKENHI Grant Number 23740162, 13J07047, and 25-7047.

Funding for the SDSS and SDSS-II has been provided by the Alfred P. Sloan Foundation, the Participating Institutions, the National Science Foundation, the U.S. Department of Energy, the National Aeronautics and Space Administration, the Japanese Monbukagakusho, the Max Planck Society, and the Higher Education Funding Council for England. The SDSS Web Site is http://www.sdss.org/.

The SDSS is managed by the Astrophysical Research Consortium for the Participating Institutions. The Participating Institutions are the American Museum of Natural History, Astrophysical Institute Potsdam, University of Basel, University of Cambridge, Case Western Reserve University, University of Chicago, Drexel University, Fermilab, the Institute for Advanced Study, the Japan Participation Group, Johns Hopkins University, the Joint Institute for Nuclear Astrophysics, the Kavli Institute for Particle Astrophysics and Cosmology, the Korean Scientist Group, the Chinese Academy of Sciences (LAMOST), Los Alamos National Laboratory, the Max-Planck-Institute for Astronomy (MPIA), the Max-Planck-Institute for Astrophysics (MPA), New Mexico State University, Ohio State University, University of Pittsburgh, University of Portsmouth, Princeton University, the United States Naval Observatory, and the University of Washington.



{\it Facilities:} \facility{Keck (DEIMOS)}, \facility{Subaru (FOCAS)}.

\end{document}